# A Model for Predicting Magnetic Particle Capture in a Microfluidic Bioseparator

**E. P. Furlani[a*], Y. Sahoo[a], K. C. Ng[a], J. C. Wortman[b], and T. E. Monk[c]**

[a]*Institute for Lasers, Photonics and Biophotonics,*

*University at Buffalo (SUNY), Buffalo, NY, 14260*

[b]*Department of Physics, Harvey Mudd College, Claremont CA, 91711*

[c]*Department of Physics, Truman State University, Kirksiville MO, 63501*

[*] E. P. Furlani is the corresponding author: email efurlani@buffalo.edu,

## Abstract

A model is presented for predicting the capture of magnetic micro/nano-particles in a bioseparation microsystem. This bioseparator consists of an array of conductive elements embedded beneath a rectangular microfluidic channel. The magnetic particles are introduced into the microchannel in solution, and are attracted and held by the magnetic force produced by the energized elements. Analytical expressions are obtained for the dominant magnetic and fluidic forces on the particles as they move through the microchannel. These expressions are included in the equations of motion, which are solved numerically to predict particle trajectories and capture time. This model is well-suited for parametric analysis of particle capture taking into account variations in particle size, material properties, applied current, microchannel dimensions, fluid properties, and flow velocity.

# 1. Introduction

In molecular biology the ability to separate biomaterials such as cells, enzymes, antigens, or DNA from their native environment is fundamental to the detection and analysis of such entities (Pankhurst et al., 2003; Molday et al., 1977). Broadly speaking, the primary function of a bioseparator is to separate a target biomaterial from a low concentration solution, and re-release it in sufficiently high concentration to enable a desired analysis. Magnetic bioseparation is a versatile and well-established method to achieve this. It involves the use of magnetic micro/nano-particles with surface treatments (immobilized affinity ligand) that are designed to bind with a target biomaterial.

Magnetic separation is usually implemented using either a direct or indirect approach (Safarýk and Safarýkova, 1977). In the more common direct approach surface-treated particles are mixed with a solution containing the target biomaterial. The mixture is allowed to incubate until a sufficient amount of biomaterial binds to the particles (bound biomaterial is said to be magnetically tagged or labeled). The labeled biomaterial is separated from the solution using a magnetic separation system, and then re-released in higher concentration for further processing.

In the indirect approach, the target biomaterial is first incubated in solution with an affinity ligand (primary antibody), which is added in free form. After a sufficient amount of biomaterial binds to the primary antibody, magnetic particles with surface-bound secondary antibodies (antibodies against the primary antibodies) are introduced, and the mixture is allowed to incubate until a sufficient amount of the target biomaterial becomes magnetically tagged. This material is then separated and re-released in higher concentration for further processing.

The use of magnetic separation in molecular biology has enjoyed a resurgence of interest over the last decade (Zborowski et al., 1995). It has advantages over competing techniques in that it is significantly faster than other methods, and enables the target biomaterial to be isolated directly from crude samples such as blood, bone marrow, and tissue homogenates. Furthermore, the relatively low permeability of the aqueous medium enables efficient coupling between the applied field and the magnetically labeled material. Moreover, since biomaterials have a relatively low intrinsic magnetic susceptibility, there is substantial contrast between tagged and untagged material, which enables a high degree of selectivity.

In conventional magnetic separation systems rare-earth magnets or electromagnets are used to produce a non-uniform field distribution throughout the separation region. When magnetic particles enter this region they experience a force that moves them towards areas of high field gradient where they can be captured. The particles have a high susceptibility and acquire a dipole moment in an external field, but quickly relax back to an unmagnetized state once the field is removed. Thus, when the field is removed the separated particles re-disperse in solution, thereby enabling the primary function of the separator, the enhancement of target material concentration.

Conventional magnetic separation systems have drawbacks in that they tend to be relatively large, costly, and complex, requiring significant energy to operate. Moreover, the accurate manipulation of microscopic particles in small sample volumes is awkward and time consuming in such systems, and the ability to precisely monitor the separation process is limited. However, advances in microsystem technology have led to the development of novel integrated magnetic bioseparation microsystems that are energy efficient and ideal for the analysis and monitoring of small samples (Gijs, 2004, Han et al., 2004). Such a system has been fabricated and characterized by Choi (Choi et al., 2000, 2001). This microsystem consists of a parallel array of rectangular conductive elements, which are semi-encapsulated in permalloy (not shown), and embedded beneath a rectangular microfluidic channel (Fig. 1A). The microsystem is small, occupying a volume of only a few cubic millimeters.

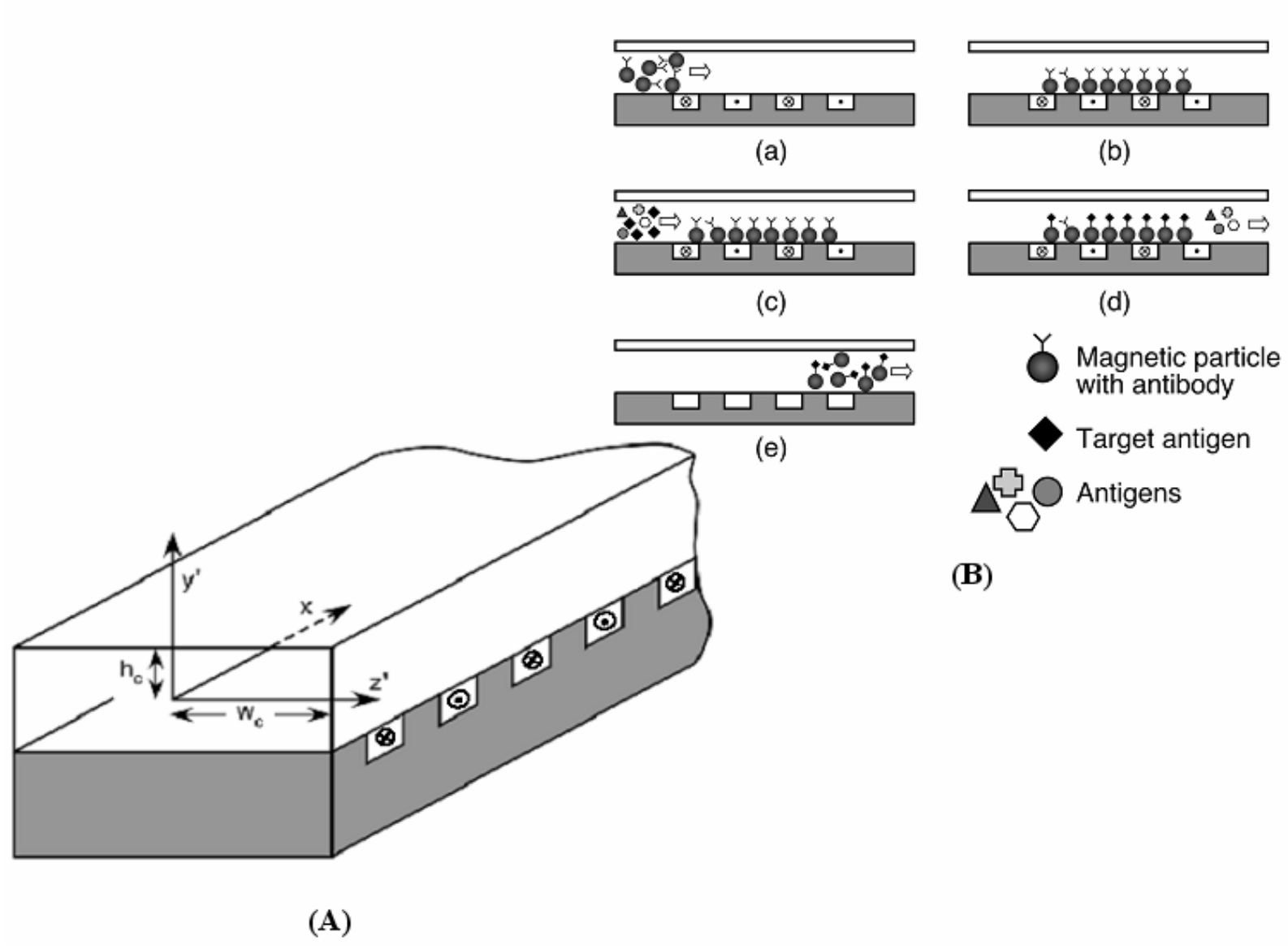

Fig.1. Microfluidic Bioseparator: (A) Perspective view with reference frame, (B) Cross-section of microfluidic bioseparator illustrating bioseparation sequence: (a) magnetic particles with surface-bound antibodies enter the microchannel, (b) energized conductive elements capture the particles, (c) target antigens are introduced into the microchannel, (d) immobilization of target antigens on magnetic particles, (e) separated material is released for further processing and analysis.

A hypothetical separation sequence for this system is depicted in a cross-sectional view in Fig. 1B. First, magnetic particles with surface-bound antibodies enter the microchannel and are captured by the energized conductive elements. Next, target antigens are introduced into the microchannel and bind to the antibodies on the captured particles, thereby becoming immobilized. Lastly, the conductive elements are de-energized and the separated material is re-released in high concentration for further processing.

In this article we develop a model for predicting the transport and capture of magnetic particles in a bioseparation microsystem similar to that developed by Choi (Choi et al., 2000, 2001). We obtain analytical expressions for the dominant magnetic and fluidic forces on the particles, and test the magnetic force expression using finite element analysis (FEA). We include the magnetic, fluidic, and gravitational forces in the equations of particle motion, and solve these equations to study particle movement within the bioseparator. Specifically, we compute particle trajectories and capture times, and show that efficient separation can be achieved in a few minutes with modest power dissipation. The model takes into account key variables including the size and magnetic properties of the particles, the dimensions and spacing of the conductive elements, the magnitude of the applied current, the dimensions of the microchannel, and the viscosity and flow rate of the fluid. An important advantage of this approach is that it is based on analytical analysis, which provides insight into the basic physics and dominant factors governing particle capture, and enables rapid parametric analysis of system performance. This is in contrast to a purely numerical approach (e.g., finite-element analysis) that tends to be awkward for parametric analysis. The model is demonstrated via application to a practical microsystem design, and our analysis indicates that efficient particle capture can be achieved in a few minutes.

## 2. Theory

In this section we derive a model for predicting the motion of a spherical magnetic particle of density $\rho_p$, radius $R_p$, volume $V_p = \frac{4}{3}\pi R_p^3$, and mass $m_p = \rho_p V_p$ in the microfluidic bioseparator shown in Fig. 1. The trajectory of the particle is governed by several forces including, (a) the magnetic force due to all field sources, (b) fluidic drag, (c) particle/fluid interactions (perturbations to the flow field), (d) inertia, (e) buoyancy, (f) gravity, (g) thermal kinetics (Brownian motion), and (h) interparticle effects that include magnetic dipole interactions. We are interested in the behavior of magnetic particles in low concentration and slow flow regimes where the magnetic and viscous drag forces dominate. Therefore, we neglect particle/fluid interactions and interparticle effects. However, we include the gravitational force, which while of second order relative to the dominant forces, is not negligible. We use classical Newtonian dynamics to study particle motion,

$$m_p \frac{d\mathbf{v}_p}{dt} = \mathbf{F}_m + \mathbf{F}_f + \mathbf{F}_g, \tag{1}$$

where $\mathbf{v}_p$ is the velocity of the particle, and $\mathbf{F}_m$, $\mathbf{F}_f$, and $\mathbf{F}_g$ are the magnetic, fluidic, and gravitational forces, respectively. The magnetic force is obtained using an "effective" dipole moment approach where the magnetized particle is replaced by an "equivalent" point dipole with a moment $\mathbf{m}_{p,eff}$ (Furlani and Ng, 2006). The force on the dipole (and hence on the particle) is given is given by:

$$\mathbf{F}_m = \mu_f \left(\mathbf{m}_{p,eff} \bullet \nabla\right) \mathbf{H}_a, \tag{2}$$

where $\mu_f$ is the permeability of the transport fluid, $\mathbf{m}_{\mathrm{p,eff}}$ is the “effective” dipole moment of the particle, and $\mathbf{H}_{\mathrm{a}}$ is the (externally) applied magnetic field intensity at the center of the particle, were the equivalent point dipole is located. If the particle is in free-space, $\mathbf{m}_{\mathrm{p,eff}} = \mathrm{V}_{\mathrm{p}}\mathbf{M}_{\mathrm{p}}$ and Eq. (2) reduces to the usual form $\mathbf{F}_{\mathrm{m}} = \mu_0 \mathrm{V}_p \left(\mathbf{M}_{\mathrm{p}} \bullet \nabla\right)\mathbf{H}_{\mathrm{a}}$, where $\mathrm{V}_{\mathrm{p}}$ and $\mathrm{M}_{\mathrm{p}}$ are the volume and magnetization of the particle, and $\mu_0 = 4\pi\times10^{-7}$ H/m is the permeability of free space.

The fluidic force is predicted using the Stokes' law for the drag on a sphere in uniform flow,

$$\mathbf{F}_{\mathrm{f}} = -6\pi\eta\mathrm{R}_{\mathrm{p}}(\mathbf{v}_{\mathrm{p}} - \mathbf{v}_{\mathrm{f}}), \tag{3}$$

where $\eta$ and $\mathbf{v}_{\mathrm{f}}$ are the viscosity and the velocity of the fluid, respectively. For the bioseparation applications of interest here, the flow in the microchannel can be considered to be laminar with a velocity profile that varies in a parabolic fashion along the height of the channel. Since the particle diameter is much smaller than the channel height, the fluid velocity is relatively constant across the particle. As such, we use Eq. (3) to estimate the drag force at a given time using the particle velocity at that time, and the fluid velocity at the position of the particle at that time. It should be noted that a rigorous analysis of the fluidic force for this application is complicated and beyond the scope of this study.

The gravitational force is given by

$$\mathbf{F}_{\mathrm{g}} = -\mathrm{V}_{\mathrm{p}}(\rho_{\mathrm{p}} - \rho_{\mathrm{f}})\mathrm{g}\,\hat{\mathbf{y}}, \tag{4}$$

where $\rho_p$ and $\rho_f$ are the densities of the particle and fluid, respectively, and $g = 9.8\ \mathrm{m/s^2}$ is the acceleration due to gravity. The gravitation force acts in the -y direction. It is worth noting that the gravitational force is often ignored when analyzing the magnetophoretic motion of submicron particles, as it is usually much weaker than the magnetic force (Furlani, 2006; Furlani and Sahoo, 2006). However, in this application the magnetic force is relatively weak and the particles can be micron-sized. Therefore, we include the gravitational force in the analysis.

We digress briefly to discuss the limitations of the Newtonian approach. As noted above, Eq. (1) does not take into account Brownian motion, which can influence particle capture when the particle diameter $D_p$ is sufficiently small. Gerber et al. have developed the following criterion to estimate this diameter (Gerber et al., 1983)

$$|F| D_p \leq kT , \tag{5}$$

where $|F|$ is the magnitude of the total force acting on the particle, k is the Boltzmann constant, and T is the absolute temperature. In order to apply Eq. (5), one needs to estimate $|F|$. If the magnetic field source is specified, one can estimate $|F|$ for a given particle by taking a spatial average of the force on the particle over the region of interest. Gerber et al. have studied the capture of $Fe_3O_4$ particles in water using a single magnetic wire, and have estimated the critical particle diameter for their application to be $D_{c,p} \equiv kT / |F|$ = 40 nm, i.e., $|F|$= 0.1 pN (Gerber et al., 1983). For particles with a diameter below $D_{c,p}$ (which is application dependent) one solves an advection-diffusion equation for the particle number density $n(\mathbf{r},t)$, rather than the Newtonian equation for the trajectory of a single particle. Specifically, $n(\mathbf{r},t)$ is governed by the following equation (Gerber et al., 1983; Fletcher, 1991),

$$\frac{\partial n(\mathbf{r},t)}{\partial t}+\nabla \bullet \mathbf{J}=0\,, \tag{6}$$

where $\mathbf{J}=\mathbf{J}_D+\mathbf{J}_A$ is the total flux of particles, which includes a contribution $\mathbf{J}_D=-D\nabla n(\mathbf{r},t)$ due to diffusion, and a contribution $\mathbf{J}_A=\mathbf{v}\,n(\mathbf{r},t)$ due to advection of particles under the influence of applied forces. Equation (6) is often written in terms of the particle volume concentration $c(\mathbf{r},t)$, which is related to the number density, $c(\mathbf{r},t)=4\pi R_p^3 n(\mathbf{r},t)/3$. The diffusion coefficient $D$ is given by the Nernst-Einstein relation $D=\mu kT$, where $\mu=1/(6\pi\eta \mathrm{R_p})$ is the mobility of a particle of radius $\mathrm{R_p}$ in a fluid with viscosity $\eta$ (Stokes' approximation). The particle drift velocity $\mathbf{v}$ in $\mathbf{J}_A$ is obtained from Eq. (1) in the limit of negligible inertia ($m_p\frac{d\mathrm{v_p}}{dt}\rightarrow 0$), i.e. by setting $\mathbf{F}_\mathrm{m}+\mathbf{F}_\mathrm{f}+\mathbf{F}_\mathrm{g}=0$. Specifically, from Eqs. (1)-(3) we find that $\mathbf{v}(\mathbf{r})=\mu\mathbf{F}(\mathbf{r})$, where $\mathbf{F}(\mathbf{r})=6\pi\eta\mathrm{R_p}\mathbf{v}_\mathrm{f}(\mathbf{r})+\mathbf{F}_\mathrm{m}(\mathbf{r})+\mathbf{F}_\mathrm{g}(\mathbf{r})$. Note that if the Stokes' drag is the only force, then $\mathbf{v}=\mathbf{v}_\mathrm{f}$. Equation (6) can be rewritten in the form,

$$\frac{\partial n(\mathbf{r},t)}{\partial t}=\frac{kT}{\left(6\pi\eta R_p\right)}\nabla^2 n(\mathbf{r},t)-\frac{1}{\left(6\pi\eta R_p\right)}\nabla\bullet\left(\mathbf{F}(\mathbf{r})n(\mathbf{r},t)\right). \tag{7}$$

In order to solve either Newton's equation (1) or the advection-diffusion equation (7), one needs an expression for $\mathbf{F}_\mathrm{f}$ and $\mathbf{F}_\mathrm{m}$, which we derive below.

### 2.1. Magnetic Force

Consider a spherical magnetic particle in the presence of an applied magnetic field $\mathbf{H}_\mathrm{a}$. Assume that particle is uniformly magnetized, and that the magnetization is a linear function of the field intensity up to saturation, at which point it remains constant at a value $M_s$. Specifically, below saturation,

$$\mathbf{M}_p = \chi_p \mathbf{H}_{in}, \tag{8}$$

where $\chi_p = \mu_p / \mu_0 - 1$ is the susceptibility of the particle, $\mu_p$ is its permeability, and $H_{in} = H_a - H_{demag}$. $H_{demag}$ is the self-demagnetization field that accounts for the magnetization of the particle, i.e. its magnetic "surface charge". It is well known that $H_{demag} = M/3$ for a uniformly polarized spherical particle with magnetization $\mathbf{M}$ in free-space (Furlani, 2001).

If the particle is suspended in a magnetically linear fluid of permeability $\mu_f$, the force on it in an applied field $\mathbf{H}_a$ is (Furlani and Ng, 2006),

$$\mathbf{F}_m = \mu_f V_p \frac{3(\chi_p - \chi_f)}{\left[(\chi_p - \chi_f) + 3(\chi_f + 1)\right]} (\mathbf{H}_a \bullet \nabla) \mathbf{H}_a. \tag{9}$$

We assume that $|\chi_f| << 1$ ($\mu_f \approx \mu_0$), in which case (9) reduces to

$$\mathbf{F}_m = \mu_0 V_p \frac{3(\chi_p - \chi_f)}{(\chi_p - \chi_f) + 3} (\mathbf{H}_a \bullet \nabla) \mathbf{H}_a. \tag{10}$$

Under this assumption we also find that

$$\mathbf{H}_{in} = \frac{3}{(\chi_p - \chi_f) + 3} \mathbf{H}_a, \tag{11}$$

and that

$$\mathbf{M}_p = \frac{3(\chi_p - \chi_f)}{(\chi_p - \chi_f) + 3} \mathbf{H}_a. \tag{12}$$

Equation (12) applies below saturation. However, when the particle is saturated, $\mathbf{M}_\mathrm{p} = \mathbf{M}_\mathrm{s}$. We account for both conditions by expressing the magnetization in terms of the applied field as follows,

$$\mathbf{M}_\mathrm{p} = f(\mathrm{H}_\mathrm{a})\mathbf{H}_\mathrm{a}, \tag{13}$$

where

$$f(\mathrm{H}_\mathrm{a}) = \begin{cases} \dfrac{3\left(\chi_\mathrm{p} - \chi_\mathrm{f}\right)}{\left(\chi_\mathrm{p} - \chi_\mathrm{f}\right) + 3} & \mathrm{H}_\mathrm{a} < \left(\dfrac{\left(\chi_\mathrm{p} - \chi_\mathrm{f}\right) + 3}{3\left(\chi_\mathrm{p} - \chi_\mathrm{f}\right)}\right)\mathrm{M}_\mathrm{s} \\ \\ \mathrm{M}_\mathbf{s} / \mathrm{H}_\mathrm{a} & \mathrm{H}_\mathrm{a} \geq \left(\dfrac{\left(\chi_\mathrm{p} - \chi_\mathrm{f}\right) + 3}{3\left(\chi_\mathrm{p} - \chi_\mathrm{f}\right)}\right)\mathrm{M}_\mathrm{s} \end{cases}, \tag{14}$$

where $\mathrm{H}_\mathrm{a} = \left|\mathbf{H}_\mathrm{a}\right|$.

The magnetic force can be decomposed into components,

$$\mathbf{F}_m(x,y) = \mathrm{F}_{mx}(x,y)\,\hat{\boldsymbol{x}} + \mathrm{F}_{my}(x,y)\,\hat{\boldsymbol{y}}, \tag{15}$$

where

$$\mathrm{F}_{mx}(x,y) = \mu_0 \mathrm{V}_\mathrm{p} f(\mathrm{H}_\mathrm{a})\left[H_{ax}(x,y)\frac{\partial H_{ax}(x,y)}{\partial x} + H_{ay}(x,y)\frac{\partial H_{ax}(x,y)}{\partial y}\right], \tag{16}$$

and

$$\mathrm{F}_{my}(x,y) = \mu_0 \mathrm{V}_\mathrm{p} f(\mathrm{H}_a)\left[H_{ax}(x,y)\frac{\partial H_{ay}(x,y)}{\partial x} + H_{ay}(x,y)\frac{\partial H_{ay}(x,y)}{\partial y}\right], \tag{17}$$

where

$$\mathbf{H}_a = \mathrm{H}_{ax}(x,y)\,\hat{\boldsymbol{x}} + \mathrm{H}_{ay}(x,y)\,\hat{\boldsymbol{y}}. \tag{18}$$

We evaluate $\mathbf{F}_m(x,y)$ in three steps. First, we obtain an expression for the field due to a single conductor. Next, we form an expression for the total field $\mathbf{H}_a$ by summing the contributions from all the conductors using superposition. Third, we substitute the expression for $\mathbf{H}_a$ into Eqs. (16) and (17), and evaluate the force.

Consider a single rectangular conductor (width 2w and height 2h) centered with respect to the origin in the x-y plane and carrying a current I into the page as shown in Fig. 2 A.

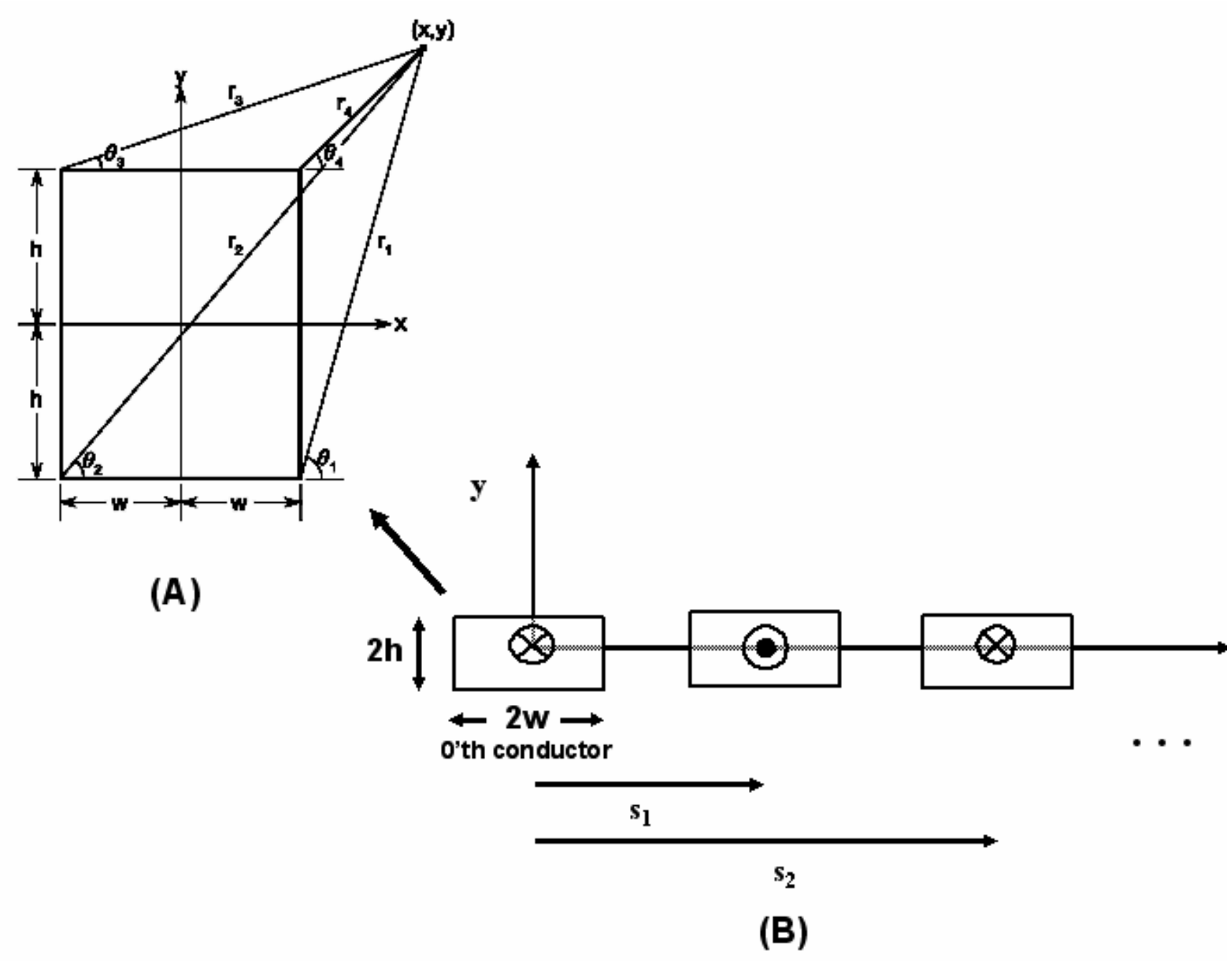

Fig. 2. Conductive elements: (A) cross-section of conductor with reference frame, (B) cross-section of array of elements.

The field components for this conductor are (Binns et al., 1992)

$$H_{ax}^{(0)}(x,y) = \frac{I}{8\pi wh}\left\{ (y+h)\big(\theta_1(x,y) - \theta_2(x,y)\big) + (x+w)\ln\left(\frac{r_2}{r_3}\right) \right. \\ \left. -(y-h)\big(\theta_4(x,y) - \theta_3(x,y)\big) - (x-w)\ln\left(\frac{r_1}{r_4}\right) \right\}, \quad (19)$$

and

$$H_{ay}^{(0)}(x,y) = -\frac{I}{8\pi wh}\left\{ (x+w)\big(\theta_2(x,y) - \theta_3(x,y)\big) + (y+h)\ln\left(\frac{r_2}{r_1}\right) \right. \\ \left. -(x-w)\big(\theta_1(x,y) - \theta_4(x,y)\big) - (y-h)\ln\left(\frac{r_3}{r_4}\right) \right\}. \quad (20)$$

where

$$r_1(x,y)=\sqrt{(x-w)^2+(y+h)^2} \qquad r_2(x,y)=\sqrt{(x+w)^2+(y+h)^2}$$
$$r_3(x,y)=\sqrt{(x+w)^2+(y-h)^2} \qquad r_4(x,y)=\sqrt{(x-w)^2+(y-h)^2} \tag{21}$$

and

$$\theta_1(x,y)=\begin{cases}\tan^{-1}\left(\dfrac{y+h}{x-w}\right) & (x>w)\\ \pi/2 & (x=w)\\ \pi+\tan^{-1}\left(\dfrac{y+h}{x-w}\right) & (x<w)\end{cases} \qquad \theta_2(x,y)=\begin{cases}\tan^{-1}\left(\dfrac{y+h}{x+w}\right) & (x>-w)\\ \pi/2 & (x=-w),\\ \pi+\tan^{-1}\left(\dfrac{y+h}{x+w}\right) & (x<-w)\end{cases} \tag{22}$$

$$\theta_3(x,y)=\begin{cases}\tan^{-1}\left(\dfrac{y-h}{x+w}\right) & (x>-w)\\ \pi/2 & (x=-w),\\ \pi+\tan^{-1}\left(\dfrac{y-h}{x+w}\right) & (x<-w)\end{cases} \qquad \theta_4(x,y)=\begin{cases}\tan^{-1}\left(\dfrac{y-h}{x-w}\right) & (x>w)\\ \pi/2 & (x=w).\\ \pi+\tan^{-1}\left(\dfrac{y-h}{x-w}\right) & (x<w)\end{cases} \tag{23}$$

Eqs. (19) and (20) are given by Binns (Binns et al., 1992), but the expression there for Eq. (20) is missing a minus sign, which has been corrected here. We substitute Eqs. (19) and (20) into Eqs. (16) and (17), and obtain the magnetic force components for a single conductor,

$$\begin{aligned}\mathrm{F}_{mx}^{(0)}(x,y)=\frac{\mu_0\mathrm{V_p}f(\mathrm{H_a})\mathrm{I}^2}{(8\pi\mathrm{wh})^2}\Bigg\{&\left[(y+h)(\theta_1-\theta_2)+(x+w)\ln\left(\frac{r_2}{r_3}\right)-(y-h)(\theta_4-\theta_3)-(x-w)\ln\left(\frac{r_1}{r_4}\right)\right]\\ &\times\left[(y+h)^2\left(\frac{r_1^2-r_2^2}{r_1^2r_2^2}\right)-(y-h)^2\left(\frac{r_4^2-r_3^2}{r_3^2r_4^2}\right)+(x+w)^2\left(\frac{r_3^2-r_2^2}{r_2^2r_3^2}\right)-(x-w)^2\left(\frac{r_4^2-r_1^2}{r_1^2r_4^2}\right)+\ln\left(\frac{r_2r_4}{r_1r_3}\right)\right]\\ &-\left[(x+w)(\theta_2-\theta_3)+(y+h)\ln\left(\frac{r_2}{r_1}\right)-(x-w)(\theta_1-\theta_4)-(y-h)\ln\left(\frac{r_3}{r_4}\right)\right]\left[(\theta_1-\theta_2)-(\theta_4-\theta_3)\right]\Bigg\},\end{aligned} \tag{24}$$

and

$$
\begin{aligned}
\mathrm{F}_{my}^{(0)}(x,y) = \frac{\mu_0 \mathrm{V_p}\, f(\mathrm{H_a})\mathrm{I}^2}{(8\pi \mathrm{wh})^2} \Bigg\{ & \left[ (y+h)(\theta_1-\theta_2) + (x+w)\ln\left(\frac{r_2}{r_3}\right) - (y-h)(\theta_4-\theta_3) - (x-w)\ln\left(\frac{r_1}{r_4}\right) \right] \\
& \times \left[ (\theta_2-\theta_3)) - (\theta_1-\theta_4) \right] \\
- \Bigg[ & (x+w)(\theta_2-\theta_3) + (y+h)\ln\left(\frac{r_2}{r_1}\right) - (x-w)(\theta_1-\theta_4) - (y-h)\ln\left(\frac{r_3}{r_4}\right) \Bigg] \\
\times \Bigg[ & (y+h)^2\left(\frac{r_1^2-r_2^2}{r_1^2 r_2^2}\right) - (y-h)^2\left(\frac{r_4^2-r_3^2}{r_3^2 r_4^2}\right) + (x+w)^2\left(\frac{r_3^2-r_2^2}{r_2^2 r_3^2}\right) - (x-w)^2\left(\frac{r_4^2-r_1^2}{r_1^2 r_4^2}\right) + \ln\left(\frac{r_2 r_4}{r_1 r_3}\right) \Bigg] \Bigg\}.
\end{aligned}
\qquad (25)
$$

Next, consider an array of $N_c$ conductors with the fist conductor centered with respect to the origin in the x-y plane, and all other conductors positioned along the x-axis as shown in Fig. 2B. The direction of current is opposite in adjacent conductors, i.e., into the page and then out of the page as shown. We identify the conductors using the index n = (0,1,2,3,4, …, $N_c$-1). The field components due to the first conductor (n = 0) are given by Eqs. (19) and (20). The n'th conductor is centered at $x = s_n$ and its field components can be written in terms of Eqs. (19) and (20) as follows,

$$
H_{ax}^{(n)}(x,y) = (-1)^n H_{ax}^{(0)}(x-s_n, y) \qquad H_{ay}^{(n)}(x,y) = (-1)^n H_{ay}^{(0)}(x-s_n, y) \qquad (n = 1,2,3,\ldots) \quad (26)
$$

The coefficient $(-1)^n$ takes into account the alternating direction of current through adjacent elements. Finally, the total field of the $N_c$ element array is obtained by summing the contributions from all the conductors,

$$
H_{ax}(x,y) = \sum_{n=0}^{N_c-1} (-1)^n H_{ax}^{(0)}(x-s_n, y), \qquad (27)
$$

$$H_{ay}(x,y)=\sum_{n=0}^{N_c-1}(-1)^n H_{ay}^{(0)}(x-s_n,y), \tag{28}$$

We substitute Eqs. (27) and (28) into Eqs. (16) and (17) and obtain the force components

$$\begin{aligned}\mathrm{F}_{mx}(x,y)=\mu_0\mathrm{V}_\mathrm{p}f(\mathrm{H}_\mathrm{a})\Bigg[&\left(\sum_{n=0}^{N_c-1}(-1)^n H_{ax}^{(0)}(x-s_n,y)\right)\left(\sum_{n=0}^{N_c-1}(-1)^n\frac{\partial H_{ax}^{(0)}(x-s_n,y)}{\partial x}\right)\\&+\left(\sum_{n=0}^{N_c-1}(-1)^n H_{ay}^{(0)}(x-s_n,y)\right)\left(\sum_{n=0}^{N_c-1}(-1)^n\frac{\partial H_{ax}^{(0)}(x-s_n,y)}{\partial y}\right)\Bigg],\end{aligned} \tag{29}$$

and

$$\begin{aligned}\mathrm{F}_{my}(x,y)=\mu_0\mathrm{V}_\mathrm{p}f(\mathrm{H}_\mathrm{a})\Bigg[&\left(\sum_{n=0}^{N_c-1}(-1)^n H_{ax}^{(0)}(x-s_n,y)\right)\left(\sum_{n=0}^{N_c-1}(-1)^n\frac{\partial H_{ay}^{(0)}(x-s_n,y)}{\partial x}\right)\\&+\left(\sum_{n=0}^{N_c-1}(-1)^n H_{ay}^{(0)}(x-s_n,y)\right)\left(\sum_{n=0}^{N_c-1}(-1)^n\frac{\partial H_{ay}^{(0)}(x-s_n,y)}{\partial y}\right)\Bigg],\end{aligned} \tag{30}$$

Equations (29) and (30) are used in the equations of motion below.

## 2.2. Fluidic Force

As noted above, we use Stokes' law to predict the fluidic drag force on the particle. Specifically, to obtain the drag force at a given time t, we substitute the particle velocity at that time $\mathbf{v}_\mathrm{p}(\mathrm{t})$, and the fluid velocity at the position of the particle at that time $\mathbf{v}_\mathrm{f}(\mathbf{x}_\mathrm{p}(\mathrm{t}))$, into Eq. (3),

$$\mathbf{F}_\mathrm{f}=-6\pi\eta\mathrm{R}_\mathrm{p}\left[\mathbf{v}_\mathrm{p}(\mathrm{t})-\mathbf{v}_\mathrm{f}(\mathbf{x}_\mathrm{p}(\mathrm{t}))\right], \tag{31}$$

To evaluate Eq. (31) we need an expression for fluid velocity $\mathbf{v}_\mathrm{f}$ in the microchannel. Let L denote the length of the channel and $h_c$ and $w_c$ denote the half-height and half-width of its rectangular cross section (Fig. 1A). The nature of the flow, laminar or turbulent, is estimated from the Reynolds number $\mathrm{Re}=\overline{\mathrm{v}}_\mathrm{f}D\rho/\eta$, where $\overline{\mathrm{v}}_\mathrm{f}$ is the average fluid velocity , D is the characteristic length of the channel (the hydraulic diameter), and $\rho$ and $\eta$ are the density and viscosity of the fluid, respectively. In bioseparation applications $\overline{\mathrm{v}}_{\mathrm{f}} < 1$ m/s, D ≈ 100 µm, $\rho \approx 1000$ kg/m$^3$, and $\eta \approx 0.001$ Ns/m$^2$. Therefore, $\mathrm{Re} < 100$ which indicates laminar flow (i.e., $\mathrm{Re} < 2300$). We assume fully developed laminar flow with the flow velocity parallel to the x-axis, and varying across the cross section,

$$\mathbf{v}_{\mathrm{f}} = \mathrm{v}_{\mathrm{f}}(y', z')\,\hat{\boldsymbol{x}}. \tag{32}$$

It is convenient to use coordinates $y'$ and $z'$ centered with respect to the cross section of the channel, and it is understood that these differ from the coordinate system used for the magnetic analysis (Fig. 2B). Here, $z'$ spans the width of the channel. The velocity profile for fully developed laminar flow is

$$\mathrm{v}_{\mathrm{f}}(y', z') = \frac{16h_c^2}{\eta\pi^3}\frac{\Delta P}{L}\sum_{n=0}^{\infty}\frac{(-1)^n}{(2n+1)^3}\left[1 - \frac{\cosh((2n+1)\pi z'/2h_c)}{\cosh((2n+1)\pi w_c/2h_c)}\right]\cos((2n+1)\pi y'/2h_c), \tag{33}$$

where $\Delta\mathrm{P}$ is the change in pressure across the length L of the channel (Ichikawa et al., 2004). The volume flow rate $Q$ through the channel is

$$Q = \mathrm{A}\overline{\mathrm{v}}_{\mathrm{f}}, \tag{34}$$

where $\mathrm{A} = 4\mathrm{h}_{\mathrm{c}} w_c$ is the cross-sectional area. If the channel is short relative to its width ($h_c / w_c \ll 1$), which is typically the case, and if we ignore the variation in velocity along the width of the channel (i.e. along the $z'$-axis), then the velocity profile reduces to,

$$\mathrm{v}_{\mathrm{f}}(y') = \frac{3\overline{\mathrm{v}}_{\mathrm{f}}}{2}\left[1 - \left(\frac{y'}{h_c}\right)^2\right]. \tag{35}$$

In order to include this expression in our analysis we rewrite it in terms of the coordinate $y$ of Fig. 2B in which $y' = y - (h + h_c + t_b)$, where $t_b$ is the thickness of the base of the channel (i.e., the distance from the top of a conductive element to the lower edge of the fluid). This gives,

$$\mathrm{v}_{\mathrm{f}}(y) = \frac{3\overline{\mathrm{v}}_{\mathrm{f}}}{2}\left[1 - \left(\frac{y - (h + h_c + t_b)}{h_c}\right)^2\right]. \tag{36}$$

Finally, we substitute Eq. (36) into Eq. (3) and obtain the fluidic force components on a particle with velocity $\mathbf{v}_{\mathrm{p}} = \mathrm{v}_{\mathrm{p,x}}\hat{\boldsymbol{x}} + \mathrm{v}_{\mathrm{p,y}}\hat{\boldsymbol{y}}$ at a position $\mathbf{x}_{\mathrm{p}} = \mathrm{x}_{\mathrm{p}}\hat{\boldsymbol{x}} + \mathrm{y}_{\mathrm{p}}\hat{\boldsymbol{y}}$,

$$\mathrm{F}_{\mathrm{fx}} = -6\pi\eta R_p\left[\mathrm{v}_{\mathrm{p,x}} - \frac{3\overline{\mathrm{v}}_{\mathrm{f}}}{2}\left[1 - \left(\frac{y_p - (h + h_c + t_b)}{h_c}\right)^2\right]\right], \tag{37}$$

and

$$\mathrm{F}_{\mathrm{fy}} = -6\pi\eta R_p\mathrm{v}_{\mathrm{p,y}}. \tag{38}$$

In these expressions, $\mathrm{v}_{\mathrm{x}}$ and $\mathrm{v}_{\mathrm{y}}$ are the components of the particle velocity. We use these in the equations of motion.

## 2.3. Equations of Motion

The equations of motion for a magnetic particle traveling through the bioseparator can be written in component form by substituting Eqs. (29), (30), (37) and (38) into Eq. (1),

$$\mathrm{m_p}\frac{d\mathrm{v_{p,x}}}{dt} = \mathrm{F_{mx}}(\mathrm{x_p}, \mathrm{y_p}) - 6\pi\eta\mathrm{R_p}\left[\mathrm{v_{p,x}} - \frac{3\overline{\mathrm{v}}_{\mathrm{f}}}{2}\left[1 - \left(\frac{\mathrm{y_p} - (\mathrm{h} + \mathrm{h_c} + \mathrm{t_b})}{\mathrm{h_c}}\right)^2\right]\right], \tag{39}$$

$$\mathrm{m_p}\frac{d\mathrm{v_{p,y}}}{dt} = \mathrm{F_{my}}(\mathrm{x_p}, \mathrm{y_p}) - 6\pi\eta\mathrm{R_p}\mathrm{v_{p,y}} - \mathrm{V_p}(\rho_{\mathrm{p}} - \rho_{\mathrm{f}})\mathrm{g}, \tag{40}$$

$$v_{p,x}(t) = \frac{dx_p}{dt}, \qquad v_{p,y}(t) = \frac{dy_p}{dt}. \tag{41}$$

Equations (39) - (41) constitute a coupled system of first-order ordinary differential equations (ODEs) that are solved subject to initial conditions for the position $x_p(0)$, $y_p(0)$, and velocity $v_{p,x}(0)$, and $v_{p,y}(0)$ of the particle. We solve these equations numerically using the Runge Kutta method.

Equations (39) - (41) predict the motion of a magnetic particle in a moving fluid that is permeated by a magnetic field. This applies to bioseparation in which the bound biomaterial is much smaller than the magnetic particle and does not appreciably influence particle motion. However, in many applications the biomaterial is much larger than a single particle. Some nominal sizes for various biomaterial are as follows (Pankhurst et al, 2003): cells (10-100μm), viruses (20-450 nm), proteins (3-50 nm), and genes (10 nm wide and 10-100 nm long). Thus, for example, if a cell is 20 microns in diameter, several micron-sized magnetic particles must bind to its surface to implement effective separation. If there are $N_p$ magnetic particles bound to a cell, the mass, volume, radius, and density of the combined cell/particle structure can be estimated using (Safarýk and Safarýkova, 1977)

$$\begin{aligned} m_{cp} &= m_{cell} + N_p\, m_p \\ &= \rho_{cell} V_{cell} + N_p\, \rho_p V_p, \end{aligned} \tag{42}$$

$$V_{cp} = V_{cell} + N_p\, V_p, \tag{43}$$

$$R_{cp} = \left( \frac{3}{4} V_{cp} \right)^{1/3}, \tag{44}$$

and

$$\rho_{cp} = \frac{\rho_{cell} V_{cell} + N_p \rho_p V_p}{V_{cell} + N_p V_p}. \tag{45}$$

The equations of motion (39) - (41) need to be modified to account these relationships. To simplify the analysis, we assume that that the magnetic force acts only on the bound magnetic particles, whereas the fluidic force acts on the entire structure, which to first order has an effective radius $R_{cp}$ . Let $x_{p,k}$ , $y_{p,k}$ denote the coordinates of the k'th magnetic particle on the surface of the cell, and let $\overline{x}_{cp}$, $\overline{y}_{cp}$ and $\overline{v}_{cp,x}$ , $\overline{v}_{cp,y}$ denote the coordinates and velocity components of the center of mass of the structure, respectively. We adapt the Eqs. (39) - (41) to the cell/particle structure and obtain,

$$m_{cp} \frac{d\overline{v}_{cp,x}}{dt} = \sum_{k=1}^{N_p} F_{mx}(x_{p,k}, y_{p,k}) - 6\pi\eta R_{cp} \left[ \overline{v}_{cp,x} - \frac{3\overline{v}_f}{2}\left[1 - \left(\frac{\overline{y}_{cp} - (h + h_c + t_b)}{h_c}\right)^2\right]\right], \tag{46}$$

$$m_{cp} \frac{d\overline{v}_{cp,y}}{dt} = \sum_{k=1}^{N_p} F_{my}(x_{p,k}, y_{p,k}) - 6\pi\eta R_{cp} \overline{v}_{cp,y} - V_{cp}(\rho_{cp} - \rho_f)g, \tag{47}$$

$$\overline{v}_{cp,x}(t) = \frac{d\overline{x}_{cp}}{dt}, \qquad \overline{v}_{cp,y}(t) = \frac{d\overline{y}_{cp}}{dt}. \tag{48}$$

In Eqs. (46) and (47) the magnetic force components $F_{mx}(x_{p,k}, y_{p,k})$ and $F_{my}(x_{p,k}, y_{p,k})$ on the k'th particle are computed using the volume and magnetic properties for that particle. In this analysis, we have assumed that the cell/particle structure is rigid, and have ignored its rotation. A more complete analysis would include the effects of rotation and structural deformation, as well as the influence of motion on the local fluid flow (i.e. the coupled structure/fluid interaction). However, these effects are beyond the scope of the present work.

## 3. Results

We use Eqs. (27)-(30) and (39)-(41) to study the bioseparator shown in Fig. 1A. We assume that the transport fluid is nonmagnetic ( $\chi_{\mathrm{f}} = 0$ ), and has a viscosity and density equal to that of water, $\eta = 0.001$ Ns/m$^2$ and $\rho_f$ = 1000 kg/m$^3$. The force calculation requires a choice of particle size and material properties. We choose a commercially available particle that is commonly used for bioapplications, the MyOne particle from Dynal Biotech (www.dynabead.com). The properties of this particle as obtained from Dynal Biotech are as follows: radius $R_p$ = 0.5 μm, density $\rho_p = 1800$ kg/m$^3$, and saturation magnetization $\mathrm{M}_s = 4.3\times10^4$ A/m. The intrinsic susceptibility $\chi_{\mathrm{p}}$ is not available. Instead, Dynal Biotech specifies an "effective" susceptibility of $\chi_{\mathrm{p,e}} = 1.4$. The values of $\chi_{\mathrm{p}}$ and $\chi_{\mathrm{p,e}}$ are measured with respect to the internal and applied field, respectively, and for a given sample the two are related as follows,

$$\chi_{\mathrm{p,e}} = \frac{\chi_{\mathrm{p}}}{1+\mathrm{N}\chi_{\mathrm{p}}}, \tag{49}$$

where N is the shape dependent demagnetization factor of the sample (e.g., N = 1/3 for a sphere). We modify our model and use $\chi_{\mathrm{p,e}}$ instead of $\chi_p$, which amounts to replacing Eq. (14) by

$$f(\mathrm{H_a}) = \begin{cases} \chi_{\mathrm{p,e}} & \mathrm{H_a} < \dfrac{\mathrm{M_s}}{\chi_{\mathrm{p,e}}} \\ \mathrm{M_s}/\mathrm{H_a} & \mathrm{H_a} \geq \dfrac{\mathrm{M_s}}{\chi_{\mathrm{p,e}}} \end{cases} . \quad (50)$$

It is important to note that commercial particles such as the MyOne are typically composites of magnetic and polymeric materials with properties that are fabrication specific, and significantly different than those of the constituent magnetic material, e.g., $Fe_3O_4$ (magnetite). Specifically, $Fe_3O_4$ has a density $\rho = 5000$ kg/m$^3$ and a saturation magnetization $\mathrm{M}_s = 4.78 \times 10^5$ A/m, both of which are substantially higher than the corresponding MyOne values. A magnetization model for $Fe_3O_4$ suitable for bioseparator analysis can be found in the literature (Furlani, 2006; Furlani and Ng, 2006 ).

We first compute the field and force on a MyOne particle due to an array of three equally spaced conductive elements. We use a reference frame centered with respect to the first element, and assign dimensions: w = 50 μm and h = 25 μm (the total width and height of the element are 100 μm and 50 μm, respectively). Each element carries a current of I = 450 mA, which corresponds to a current density of $J = 9\times10^7$ A/m$^2$. The direction of current alternates from element to element as shown in Figs. 3c and 4c. The first element is centered at the origin of the x-y plane with its current directed into the page, and the elements are spaced 100 μm apart (edge to edge). The field and force components are evaluated along a horizontal line $-2w \le x \le 10w$ at y = h +10 μm (10 μm above the top of the elements), and are shown in Figs. 3 and 4, respectively, along with corresponding data obtained using finite element analysis (FEA). The FEMLAB program from COMSOL was used for the FEA. Note that $B_x$ peaks at the center of each element, but alternates in sign from one element to the next due to the alternating current. The components $B_y$ , $F_{mx}$ and $F_{my}$ peak near the edges.

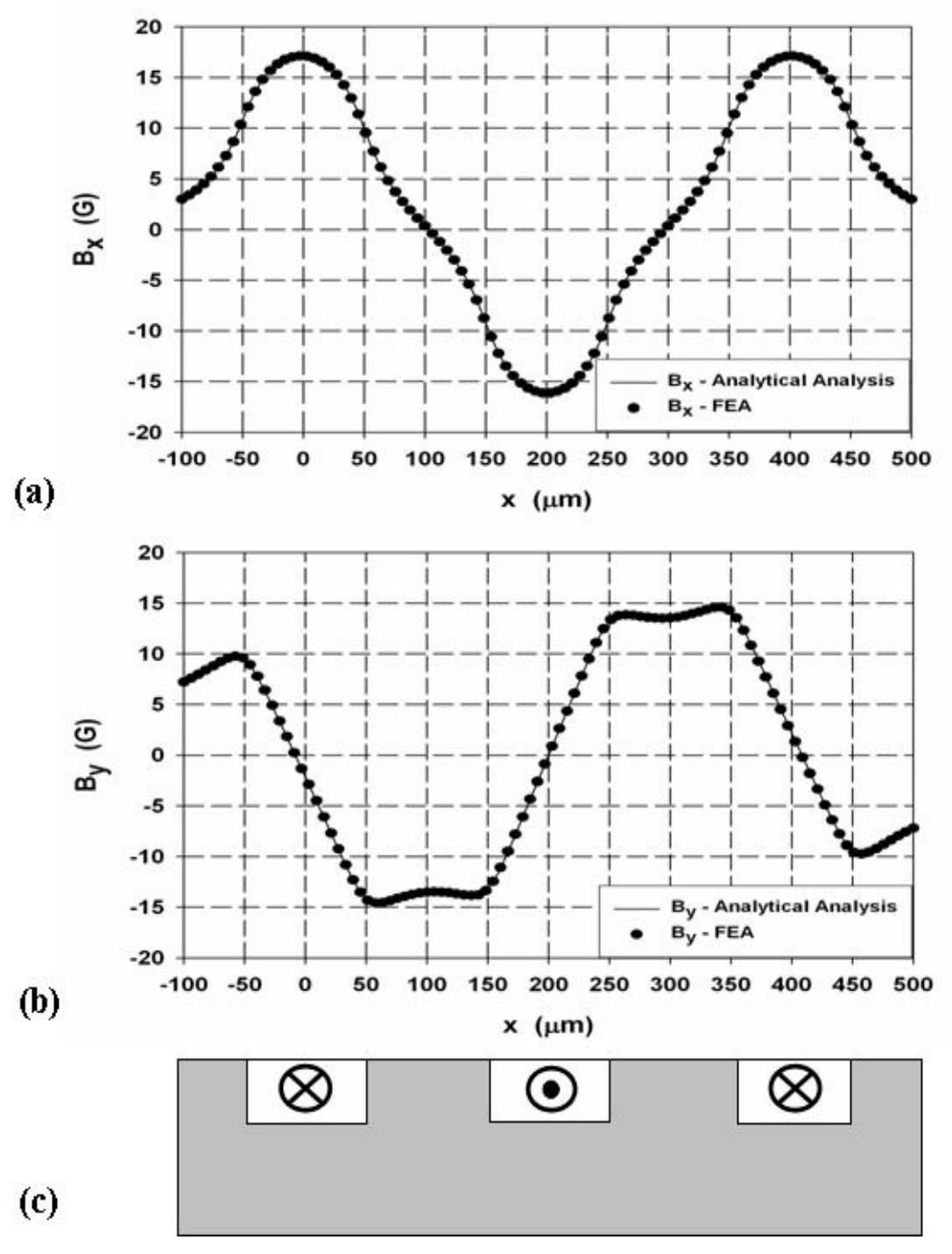

Fig. 3. Magnetic field on a MyOne particle along a horizontal line 10 Pm above an array of three conductive elements: (a) Bx, (b) By, and (c) cross-section showing of three element array embedded in the substrate ( x = FEA).

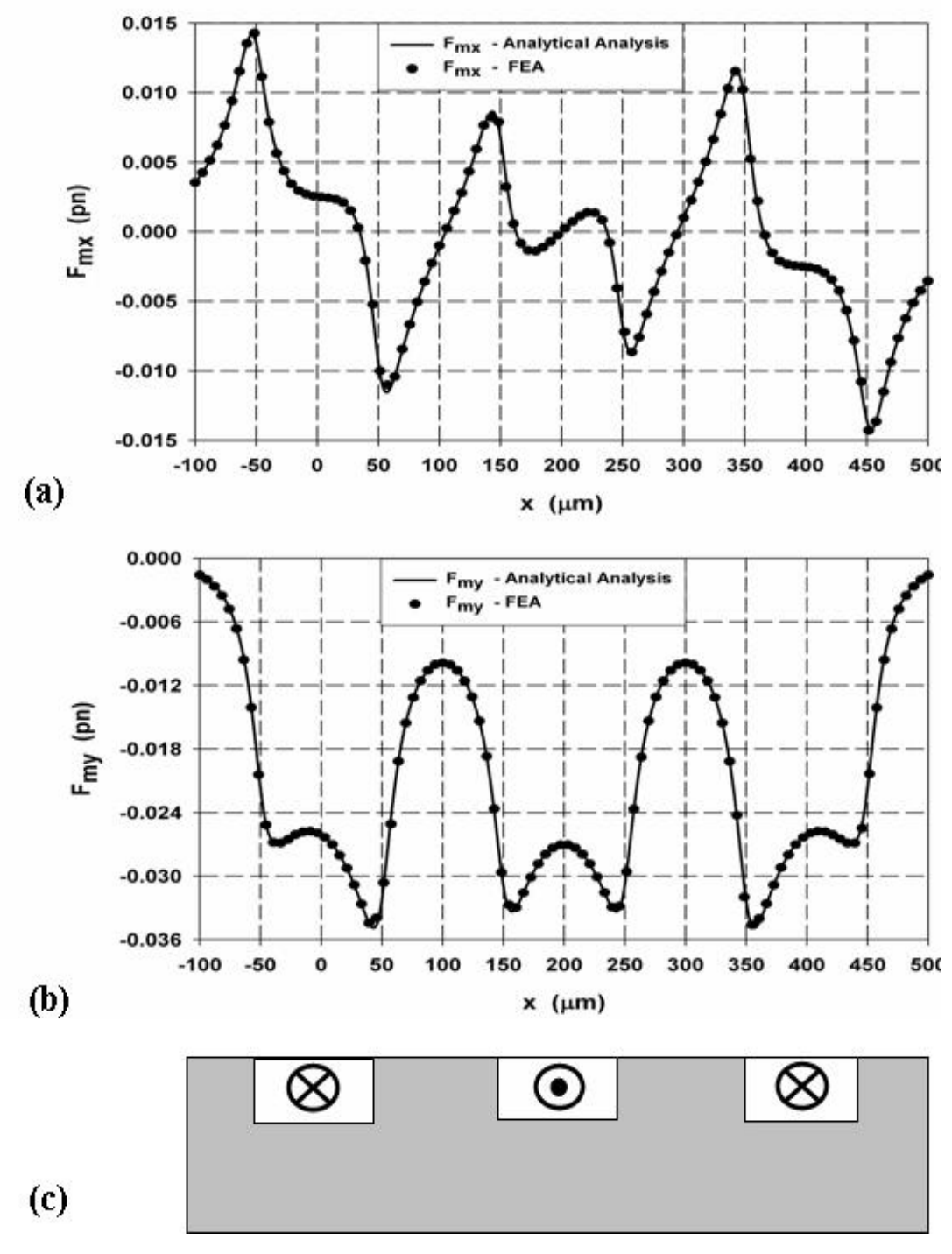

Fig. 4. Magnetic force on a MyOne particle along a horizontal line 10 Pm above an array of three conductive elements: (a) Fmx, (b) Fmy, and (c) cross section showing of three element array embedded in the substrate ( x = FEA).

Next, we study the fluidic force on the particle. The total height of the microchannel is set to 100 μm ($h_c = 50\ \mu m$), and we assume that the average fluid velocity is $\overline{v}_f = 0.1$ mm/s. We use Eq. (37) and compute $F_{fx}$ along a vertical line $h \leq y \leq h + 2h_c$ (i.e., from the bottom to the top of the channel, $t_b = 0$ in this analysis). The force profile, which is plotted in Fig. 5, has a parabolic shape with a maximum of approximately $F_{fx}$ = 1.4 pN midway in the channel and two minima of zero at either edge. Thus, the magnetic force at the top of a conductive element (near its edge) will hold a particle in place once it reaches the bottom of the channel, as the fluidic force is essentially zero there.

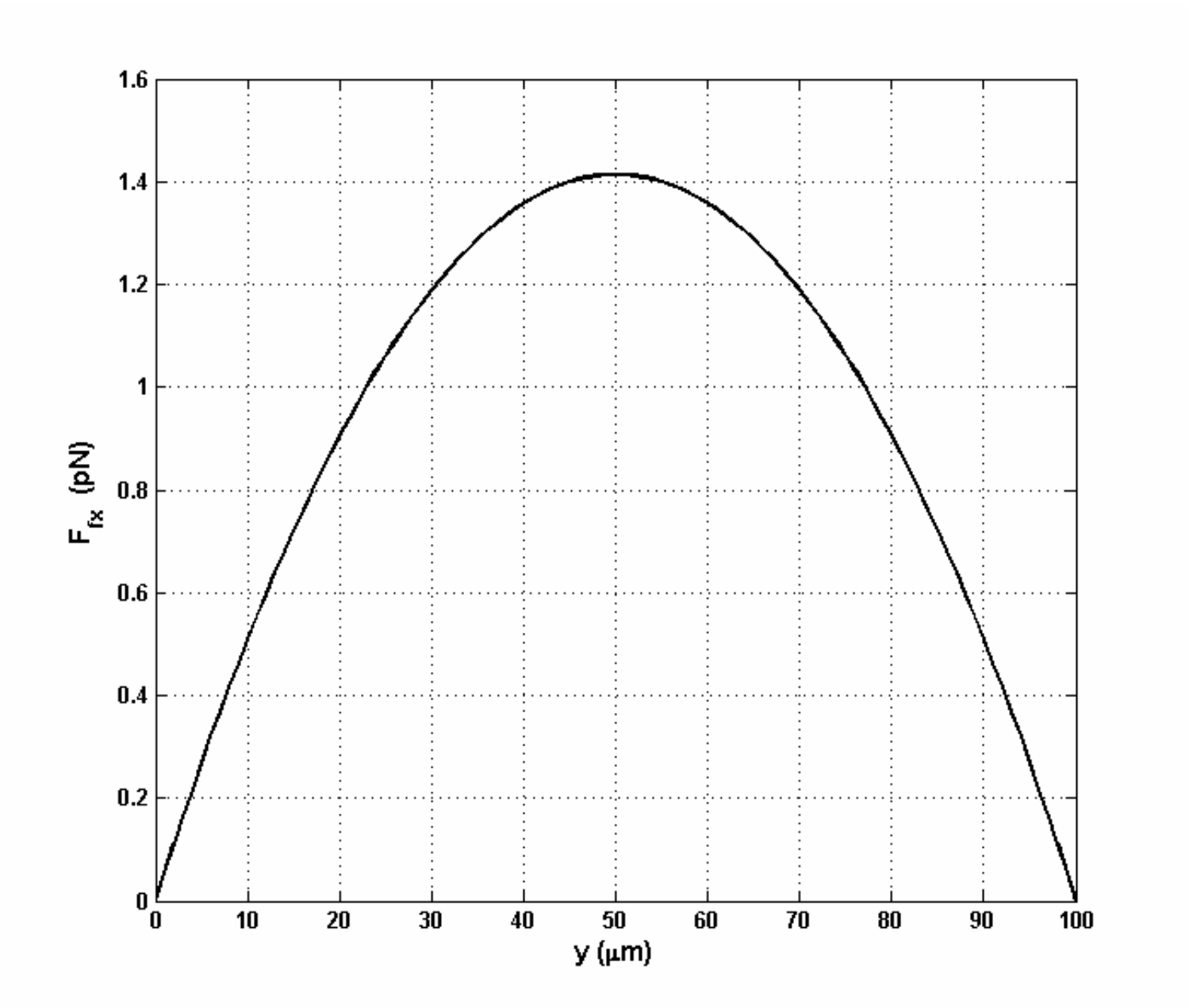

Fig. 5. Horizontal component of the fluidic force along a vertical line that spans the height of the microchannel.

We now study the dynamics of an isolated particle as it moves through the bioseparator. Again, the total height of the microchannel is 100 μm, and we assume that there is an array of 40 identical conductive elements immediately beneath its base ($t_b = 0$). Each element has dimensions w = 50 μm and h = 25 μm and carries a current I = 450 mA, which alternates in direction from one element to the next as shown in Fig. 1A. The elements are spaced 100 μm apart (edge to edge). As above, we use a reference frame centered with respect to the first element, and therefore the last element of the array ends at x = 7850 μm. We examine the trajectory of the particle as a function of its vertical entry point into the bioseparator. Specifically, we assume that the particle enters to the left of the first element at $x(0) = -5w$, and compute its trajectory as a function of its initial height above the conductive elements: Δy = 10 μm, 20 μm, …, 90 μm (i.e., initial heights of y(0) = 35 μm, 45 μm, …, 115 μm). The average fluid velocity is $\overline{v}_f = 0.1$ mm/s, and the particle enters the channel with this initial velocity, $v_{p,x}(0) = 0.1$ mm/s. The computed particle trajectories are shown in Fig. 6. It is easy to identify each trajectory with the corresponding entry height into the microchannel, as this height is the starting point of the trajectory on the y-axis. For this analysis we integrated Eqs. (39) and (40) using the fourth-order Runge-Kutta method, and it took less than one minute to complete the simulation. The results indicate that the bioseparator will capture particles that enter the microchannel 0-80 μm above the conductive elements. However, if its entry point is higher than this, a particle will pass through the microchannel (beyond the array which ends at x = 7850 μm). Note that the trajectory plots are slightly irregular because the x-component of the magnetic force reverses sign each time the particle passes over a conductive element as shown in Fig. 4a. Thus, the particle experiences acceleration followed by deceleration as it passes over an element, which gives rise to the irregular plots.

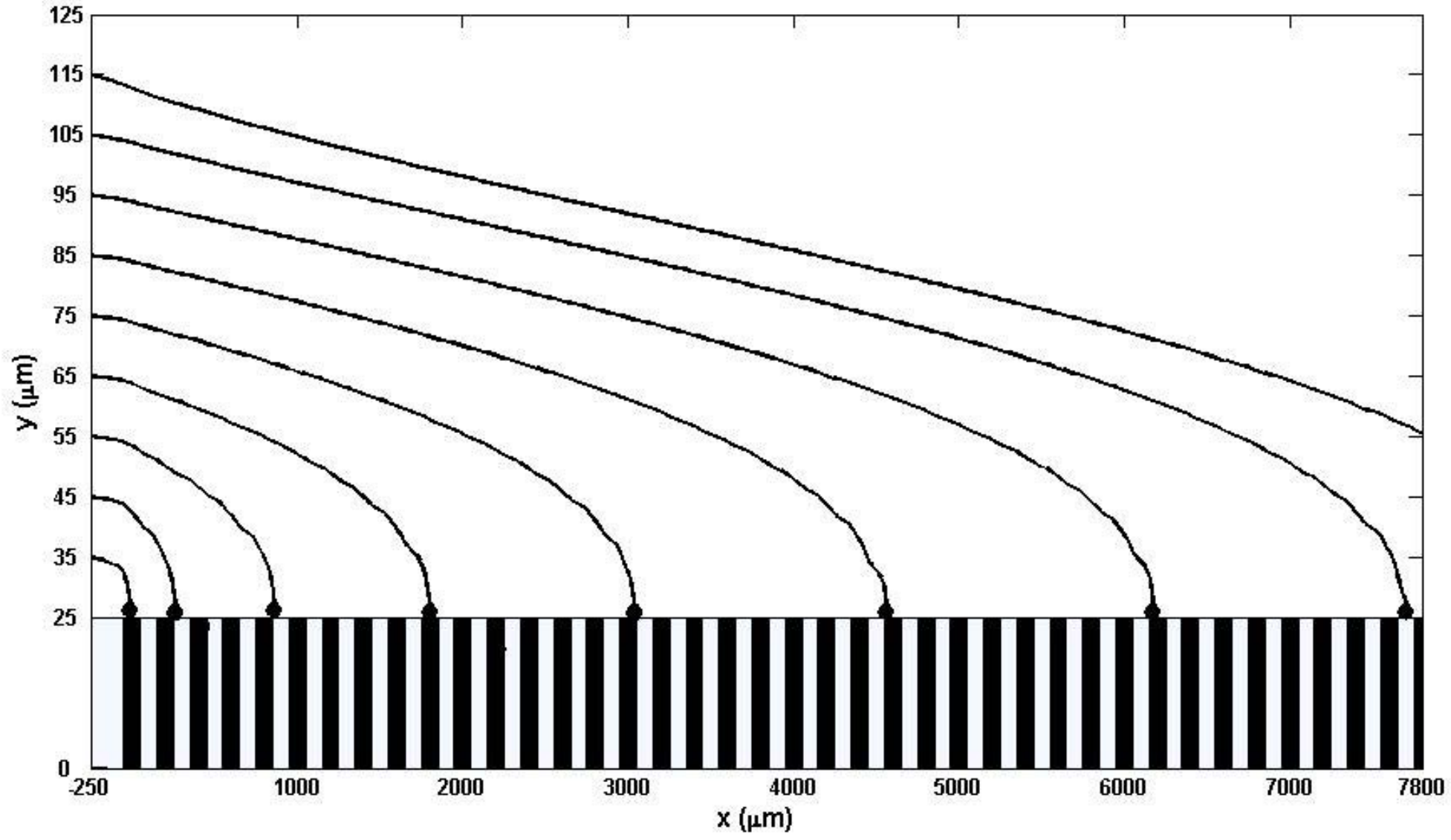

Fig. 6. Particle trajectories vs. initial entry height above conductive elements.

The capture time, i.e., the time it takes for the particle to reach the bottom of the microchannel where it will be held in place, is plotted as a function of the entry height in Fig. 7.

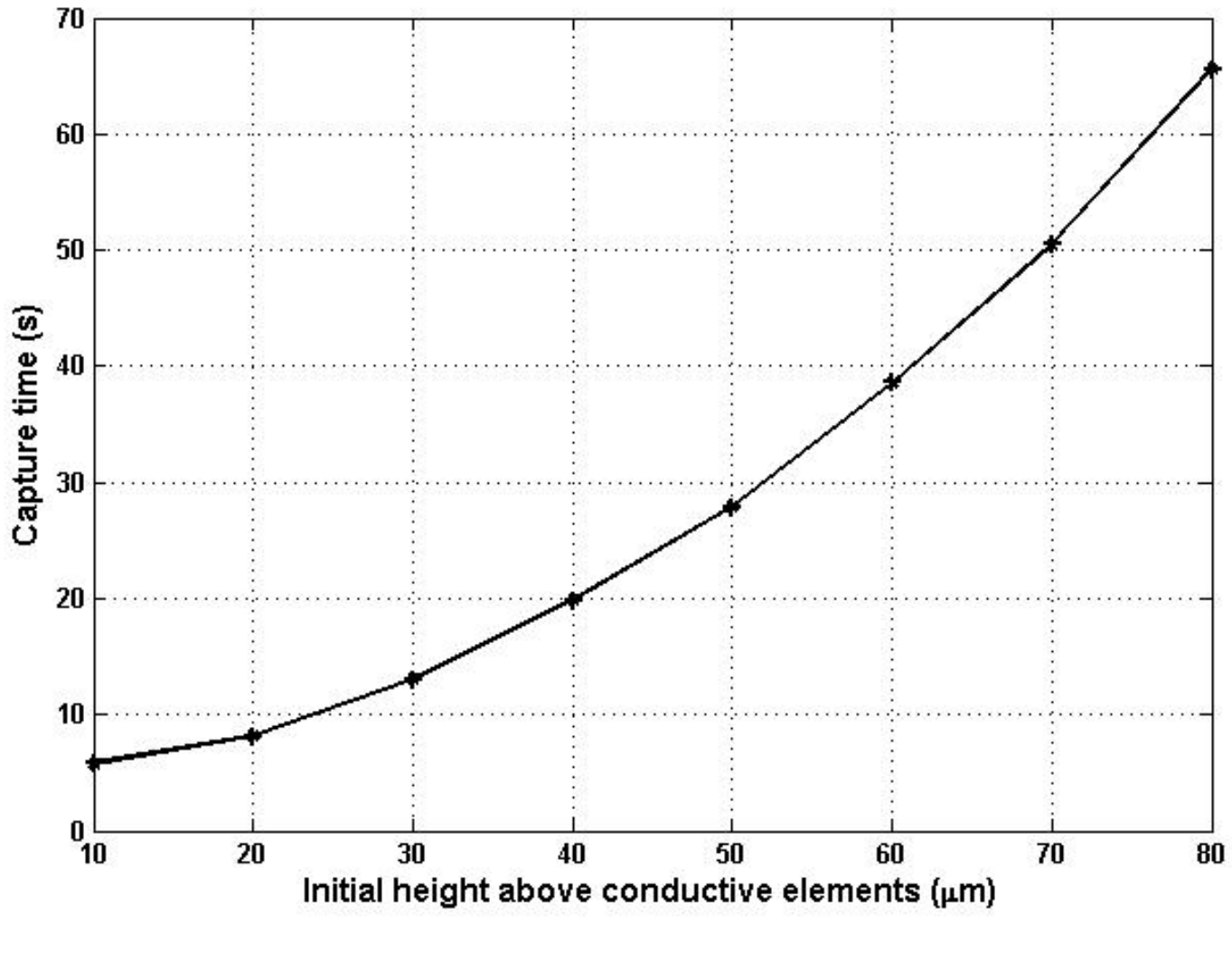

Fig. 7. Capture time vs. initial entry height above conductive elements.

Particles that enter the microchannel 0-80 μm above the conductive elements will be captured within 70 s.

It is instructive to note that the gravitational force (including buoyancy) on a MyOne particle is $F_g = 0.0041$ pN. This is much smaller than the magnetic force $F_{my} = 0.034$ pN that it experiences 10 um above the edge of an element (Fig. 4b). Now, if $F_g$ were the only force acting on the particle, it would obtain a downward vertical drift velocity of 0.435 μm/s, which follows from a balance between the gravitational and fluidic drag force. Thus, it would take 161 seconds for a MyOne particle to reach the bottom of the microchannel starting from a height of 70 μm above the elements. This is much longer than the 50 seconds it takes to travel the same distance under the combined influence of $F_{my}$ and $F_g$ (Fig. 7). Thus, the magnetic force dominates the gravitational force. However, $F_g$ needs to be included in the analysis because its contribution to the particle motion, while small, is not negligible.

Lastly, we consider the power dissipated by the bioseparator. If the conductive elements (which extend beyond the width of the microchannel) are 1 mm long and made of copper ($\sigma = 5.8\times10^7$ S/m), then each element will have a resistance of 3.45 mΩ. A potential difference of 1.55 mV must be applied across the length of each element to produce the specified current of $I = 450$ mA. If there are 40 elements, and if the elements are electrically connected in a serpentine fashion, the then the power dissipated by the microsystem will be approximately 36.4 mW, wherein 28 mW will be dissipated in the elements, and 8.4 mW in the connectors. Here, we assume that the conductive elements extend far enough beyond the microchannel so that the magnetic field due to the connectors does not appreciably alter the capture field. This is more of a design issue, but it is reasonable to assume that the connectors should be spaced from the microchannel a distance at least equal to the gap between neighboring elements.

## 4. Conclusion

We have presented a model for predicting particle capture in a magnetophoretic microfuidic bioseparator. The model is based on analytical analysis and is well-suited for rapid parametric studies of particle trajectory and capture time as a function of key variables including the size and properties of a particle, its entry point into the microfluidic system, the dimensions and spacing of the conductive elements, the applied current, the dimensions of the microchannel, and the viscosity and flow rate of the fluid. We have applied the model to a bioseparator design, and our results indicate that efficient particle capture can be achieved within a few minutes with only modest energy dissipation.